\documentclass[prd,superscriptaddress,superscriptaddress,%
twocolumn,showpacs,preprintnumbers,amsmath,amssymb]{revtex4-1}
\usepackage{graphicx}
\usepackage[breaklinks, colorlinks=true, pdfstartview=FitV,%
 linkcolor=red, citecolor=blue, urlcolor=blue]{hyperref}
\usepackage{color}

\newcommand{\fm}{\text{fm}}
\newcommand{\calB}{\mathcal{B}}
\newcommand{\calE}{\mathcal{E}}

\newcommand{\GeV}{\,\text{GeV}}

\newcommand{\xt}{\boldsymbol{x}_\perp}

\newcommand{\tr}{\text{tr}}
\newcommand{\Nc}{N_{\text{c}}}

\newcommand{\Pt}{P_{\text{T}}}
\newcommand{\Pl}{P_{\text{L}}}

\begin{document}

\title{Turbulent pattern formation and diffusion
       in the early-time dynamics\\
       in the relativistic heavy-ion collision}
\author{Kenji Fukushima}
\affiliation{Department of Physics, Keio University,
             Kanagawa 223-8522, Japan}

\begin{abstract}
  We propose a picture of turbulent pattern formation in the
  relativistic heavy-ion collision, which follows an efficient
  process to break color strings and dispose energy in the whole phase
  space.  We perform numerical simulations using the SU(2) pure
  Yang-Mills theory in a non-expanding box to observe a dynamical
  phenomenon in the transverse plane akin to the domain growth in
  time-dependent spin systems.
\end{abstract}
\maketitle


\section{Introduction}

Relativistic heavy-ion collision experiments at RHIC in BNL and at LHC
in CERN have successfully created a quark-gluon plasma and probed into
its detailed properties;  among milestones in the quark-gluon plasma
research, the recognition of the so-called \textit{perfect fluidity},
i.e.\ smallness of the shear viscosity to the entropy density ratio
was the most influential one, which has triggered interdisciplinary
discussions over many fields including nuclear physics and
super-string theories.

Thanks to the tremendous developments of the lattice simulation of
quantum chromodynamics (QCD)~\cite{Philipsen:2012nu} and the
(dissipative) hydrodynamic model~\cite{Hirano:2012kj}, we have reached
a reasonable understanding on the static properties of high-$T$ QCD
matter and the subsequent dynamical evolution.  It is, however, still
far from establishing a firm theoretical framework for the
pre-thermalization stage.  Generally speaking, the thermalization
process out of equilibrium is a ubiquitous but complicated problem,
and theoretical research mostly relies on numerical methods.

Luckily, in the case of the relativistic heavy-ion collision at high
enough energy, the very early-time dynamics at the time scale of order
of $\tau\sim Q_s \lesssim 0.1\fm/c$ can be expressed in terms of
coherent gluon fields, where $Q_s$ is the saturation
momentum~\cite{Iancu:2002xk}.  The initial state characterized by
$Q_s$ is sometimes referred to as the ``glasma'' initial
condition~\cite{Lappi:2006fp}.  In this glasma picture the most
important is the presence of boost-invariant longitudinal
chromo-electric and chromo-magnetic fields, $\calE^\eta$ and
$\calB^\eta$, the intensity of which is given by $Q_s$ again.

In contrast to the glasma stage, the time scale when the hydrodynamic
model starts working is of order $\tau\lesssim 1\fm/c$.  It is an
urgent theoretical problem to fill in the gap between these times
scales.  Along this line there are many theoretical attempts based on
the glasma simulation~\cite{Romatschke:2005pm,Fujii:2008dd,%
Fukushima:2011nq}, the plasma instability~\cite{Mrowczynski:1993qm},
the hard-loop expansion~\cite{Mrowczynski:2004kv}, the kinetic
description~\cite{Bratkovskaya:2000qy}, the holographic
duals~\cite{Janik:2006gp,Chesler:2008hg}, and the classical Yang-Mills
simulations~\cite{Muller:1992iw,Berges:2007re}.

In this work we solve the Yang-Mills equation of motions starting with
the glasma initial condition.  The question is then; what is the most
likely candidate for the mechanism to ``decohere'' the longitudinal
$\calE^\eta$ and $\calB^\eta$ in such a short time scale.  This kind
of decoherence problem is a quite generic problem we may encounter in
various circumstances (see e.g.\ \cite{Dusling:2010rm} for a
scalar-model study).  Our proposal is that the turbulent diffusion
should be the driving force for this;  indeed it is known in many
physical phenomena that the turbulence is a much faster process than
the typical molecular diffusion by several orders of magnitude.

In the context of the RHIC and LHC physics, the role of the turbulence
has been emphasized as a possible account for the smallness of the
viscosity to the entropy density ratio~\cite{Asakawa:2006tc} --
because the energy transport goes efficiently, an anomalously small
viscosity arises generally in a turbulent flow.  The actual
calculation assumes a random background distribution of
chromo-fields~\cite{Kirakosyan:2012aq}.  Therefore, we still need to
consider from where these fields are generated, and the glasma
simulation is indispensable to answer such a question.  The
turbulence, especially the wave turbulence, has also been investigated
numerically and
analytically~\cite{Berges:2008mr,Schlichting:2012es,Kurkela:2012hp,%
Berges:2013eia}.  It has been understood that the Kolmogorov-type
cascade leads to a power-law spectrum (where the power index may take
different values at strong coupling).  The Kolmogorov behavior is,
however, realized in a system with a well-developed inertial
region~\cite{Zakharov}.  This means that we have to wait for a certain
time until the power-law spectrum grows steadily, while what we want
to clarify is not the steadiness but the rapid reorganization from the
initial state.  We should, hence, disturb the initial system with
substantially large fluctuations that breach the boost invariance.

For this purpose, in this work, we turn off the effect of the
expanding geometry.  We do this because the expanding geometry is
singular at the initial time and it is difficult to disturb the
initial state without ambiguity.  Besides, the expansion quickly
renders the transverse dynamics frozen, and thus, one should carefully
formulate the initial spectral shape (involving the UV divergence) and
also elaborate the proper renormalization (or subtraction)
procedures~\cite{Dusling:2010rm,Fukushima:2006ax}.  Otherwise, useful
information on underlying physics can be easily diluted and even
concealed by the effect of expansion.  These are not simply technical
problems;  the expanding geometry represents curved space-time and
quantum fluctuations on such curved space-time are distorted, so that
the physical vacuum should be Bogoliubov transformed from the vacuum
in flat space-time.  Interestingly enough, as we will find later, a
particular initial condition corresponding to the heavy-ion collision
already captures the essential features of the anisotropic dynamics.
Moreover, we have checked whether we can confirm the same observation
using a code for the expanding case, and have found similar behavior
if we employ large fluctuations, while the non-expanding simulation
always leads to inhomogeneous pattern formation for any (small)
fluctuations.

One might feel that the approximation to adopt a non-expanding box be
artificial simplification.  However, it is just obvious that the
expansion effects should delay the decohering processes, and
therefore, one must first understand the decohering mechanism for the
non-expanding system;  otherwise it is impossible to give any account
for the expanding case.  Then, one should test the presumed mechanism
to see whether it works to overcome the delay in the expanding case.
This is why we dare to drop the expansion off;  the relevance to the
experiment becomes less, but the theory becomes more well-behaved
thanks to this simplification.

The most important point in this work is that the longitudinal and the
transverse dynamics behave totally differently at early time when the
anisotropy from the collision geometry is huge.  One can presume by
intuition that the longitudinal decoherence should go much faster than
the transverse one;  otherwise the isotropization is never achieved.
Indeed we will confirm this anticipation, and find that it indeed
happens in a very interesting manner.


\section{Formulation}

Most importantly, we can make use of the glasma initial
condition~\cite{Kovner:1995ja,Krasnitz:1998ns} as it is even in the
non-expanding case because the initial fields lie only on the
transverse plane.  In terms of the link variables on the
lattice, the canonical momenta leads to the following time evolution,
\begin{equation}
 U_i(t+2\Delta t) = \exp\bigl[ -i g E^i(t+\Delta t)
  \cdot 2\Delta t \bigr] U_i(t) \;,
\end{equation}
where the time arguments are shifted in accord to the leapfrog
algorithm which preserves the Gauss law exactly.  We omit writing the
lattice spacing $a$ throughout this paper.  The classical Yang-Mills
equations of motion (Hamilton's equations) read,
\begin{equation}
 \begin{split}
 & E^i(t+\Delta t) - E^i(t-\Delta t) \\
 &\qquad = 2\Delta t \cdot \frac{i}{2g}
  \sum_{j\neq i}\bigl[ U_{ji}(t) + U_{-ji}(t) - \text{(h.c.)} \bigr]
 \end{split}
\end{equation}
in the temporal axial gauge; $U_t=1$.  The initial condition is given
in a standard way, which simplifies particularly for the SU(2) color
group~\cite{Krasnitz:1998ns} as
\begin{align}
 U_i &= \bigl( U_i^{(1)}+U_i^{(2)} \bigr)\bigl( U_i^{(1)\dagger}
  +U_i^{(2)\dagger} \bigr)^{-1} \;,
\label{eq:ini1}\\
 E^z &= \frac{-i}{4g}\sum_{i=x,y} \Bigl\{ (U_i\!-\!1)\bigl(
  U_i^{(2)\dagger} \!+\! U_i^{(1)\dagger} \bigr)
   + \bigl[ U_i^\dagger(x \!-\! \Delta x_i) \!-\! 1 \bigr] \notag\\
 & \quad\times\bigl[ U_i^{(2)\dagger}(x\!-\!\Delta x_i)
          \!-\! U_i^{(1)\dagger}(x\!-\!\Delta x_i) \bigr]
   - \text{(h.c.)} \Bigr\}
\label{eq:ini2}
\end{align}
with the pure gauge configurations,
$U_i^{(m)}(\xt)=V^{(m)}(\xt)V^{(m)\dagger}(\xt\!+\!\Delta x_i)$,
and the gauge rotation, $V^{(m)\dagger}=e^{ig\Lambda^{(m)}}$, by the
static potential obtained as a solution of the Poisson equation,
$\partial_\perp^2\Lambda^{(m)} = -\rho^{(m)}$.

We assume the Gaussian distribution for the color source;
$\langle\rho^{(n)}(\xt)\rho^{(m)}(\xt')\rangle \!=\! \delta^{nm}
g^2\mu^2 \delta(\xt\!-\!\xt')$, where $\mu$ is supposed to be related
to the characteristic scale $Q_s$ as was mentioned in the previous
section.  If we solved the time evolution with the initial
condition~\eqref{eq:ini1} and \eqref{eq:ini2}, there would appear no
dependence on the longitudinal coordinate (i.e.\ $z$ in the present
case without expansion, corresponding to $\eta$ in the Bjorken
coordinates).  This means that QCD color strings extend along the
$z$-direction at initial time.  We shall introduce a minimal
perturbation to make it the clearest how these strings are disrupted
by extra fluctuations;
\begin{equation}
 E^i = g^3\mu^2 [f(z\!-\!\Delta z)-f(z)] \xi^i \;,~~
 f(z)=\Delta\cos(2\pi z/L_z) \;,
\label{eq:fluc}
\end{equation}
where $\langle\xi^i(\xt)\xi^j(\xt')\rangle=\delta^{ij}
\delta^{(2)}(\xt\!-\!\xt')$ and $\delta E^\eta$ solved from the Gauss
law.  In this way we put a seed of electric-field amplitude
$\propto \Delta$ at the lowest non-zero momentum
$|k_z^{\text{(min)}}|=2\pi/L_z$.  As long as the instability stays
weak, the linear superposition gives a good approximation for the
results with more general fluctuations~\cite{Fukushima:2011nq}.
In this work, however, we will also choose a non-small $\Delta$ to
test the robustness of what we discover.  In principle, we should
generate the quantum fluctuations according to the ground state
(Gaussian) wave-function, and take the ensemble average over all
fluctuations, which would lead to the UV divergence of the zero-point
oscillation energy.  To avoid this complication, in this work, we pick
up a ``representative'' of the configuration by Eq.~\eqref{eq:fluc}.
This simplification would affect quantitative details such as the
precise time scale of the decoherence, but should be harmless to the
qualitative nature of the phenomenon that we will discuss.


\section{Numerical Results}

First let us address the case without $z$-dependent fluctuations.
With unbroken translational invariance in the $z$-direction, the
transverse pressure $\Pt$ approaches a finite value, while the
longitudinal pressure $\Pl$ decreases to vanish asymptotically,
where they are, respectively, defined as
\begin{align}
 & \Pl = \tr\bigl[ (E^x)^2 \!+\! (E^y)^2 \!-\! (E^z)^2
         \!+\! (B^x)^2 \!+\! (B^y)^2 \!-\! (B^z)^2 \bigr] \;,
 \notag\\
 & \Pt = \tr\bigl[ (E^z)^2 + (B^z)^2 \bigr] \;.
\end{align}
In our numerical computation we use $g^2\mu=120/L_\perp$
(corresponding to the choice $g^2\mu\sim 2\GeV$) with $g=2$ and the
transverse and longitudinal site numbers, $L_\perp=L_z=96$.  We note
that the initial energy density is both UV and IR
singular~\cite{Kovchegov:2005ss}:
\begin{equation}
 \varepsilon(t=0) = \Nc (\Nc^2-1) \frac{(g^2\mu)^4}{8\pi^2 g^2} \biggl[
  \ln \frac{\Lambda_{\rm UV}}{m_{\rm IR}} \biggr]^2 \;,
\end{equation}
where $\Lambda_{\rm UV}$ and $m_{\rm IR}$ are UV and IR cutoff scales,
respectively.  This singularity is problematic in a non-expanding box,
while the time evolution soon diminishes this singularity in the
expanding case~\cite{Kovchegov:2005ss}.  In our numerical simulation,
thus, we need to introduce a UV cutoff $(k_\perp)_{\rm max}$ when we
solve the Poisson equation, i.e.\ higher modes with
$k_\perp > (k_\perp)_{\rm max} = 32\cdot 2\pi/L_\perp \sim 1.7 g^2\mu \sim 3.4\GeV$
are dropped to get the results presented in this paper.  We have then
confirmed that our results have only minor dependence on $L_\perp$ as
long as we keep the same $(k_\perp)_{\rm max}$.  We also note that,
because of the color string, the initial $\Pl$ starts from a negative
value (i.e.\ two nuclei feel an attractive force).

It is already a non-trivial observation that $\Pl$ vanishes at late
time.  In the expanding case, since the system is stretched and
diluted, one may anticipate $\Pl\to0$ as a result of the free
streaming.  In the present simulation, however, the box does not
expand and nothing streams out, so that $\Pl\to0$ is purely realized
by the choice of the initial conditions~\eqref{eq:ini1} and
\eqref{eq:ini2}.  This implies that $\Pl\to0$ even in the expanding
glasma should be attributed to not the expansion but the initial
conditions.  In other words, the free streaming is not the reason, but
the physical interpretation of the result.

\begin{figure}
 \includegraphics[width=\columnwidth]{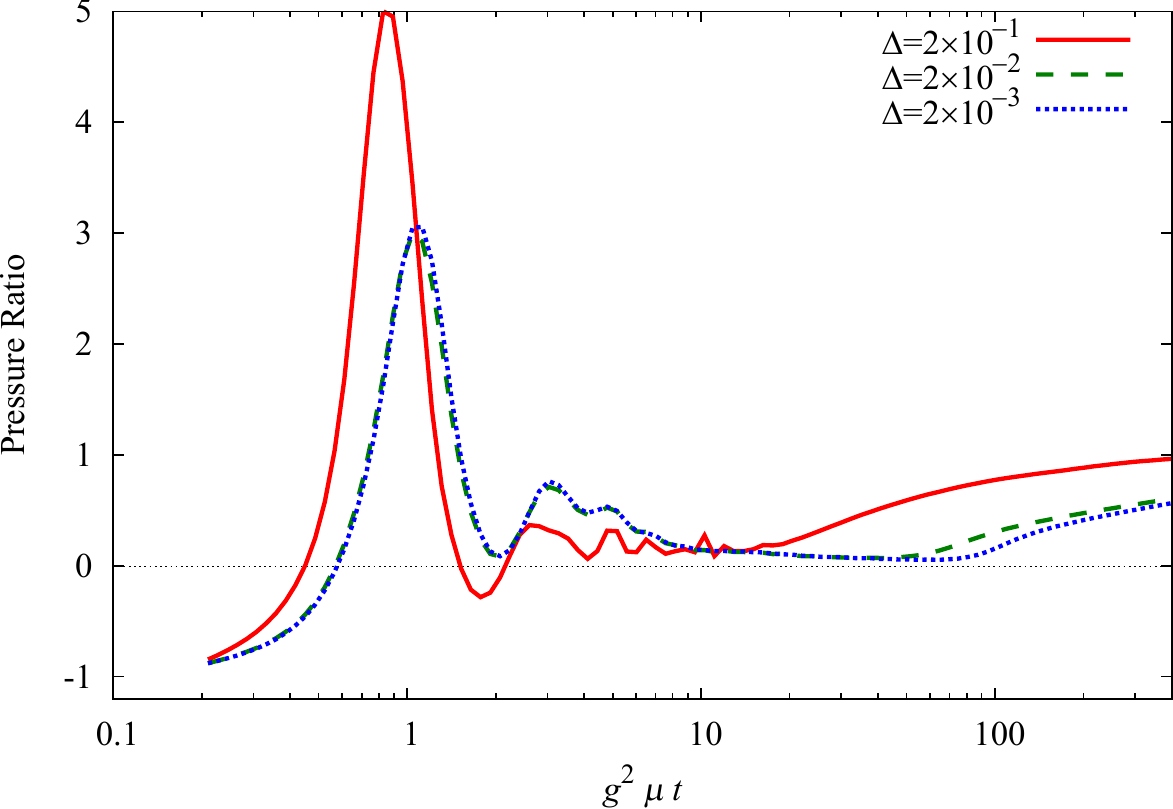}
 \caption{Pressure ratio $\Pl/\Pt$ as a function of dimensionless
   time.  Without fluctuation the ratio approaches zero, while it goes
   to a non-zero constant if fluctuations are implemented.  An
   ensemble average is taken over 50 configurations.}
 \label{fig:pressure}
\end{figure}

We shall next proceed to the results with $z$-disturbing fluctuations,
as shown in Fig.~\ref{fig:pressure}.  We here adopt three different
$\Delta$;  a substantially large $\Delta = 0.2$ gives the energy
density from fluctuations of the same order of magnitude as the
background fields.  Therefore, this value is a kind of upper bound
above which solving the classical equations of motion is no longer
justified.  A marginal $\Delta=0.02$ is much safer;  the initial
energy density is dominated by the background fields, and the time
evolution is almost identical with the case with even smaller
$\Delta=0.002$, as is manifested in Fig.~\ref{fig:pressure}.  (To
avoid making the figure too busy, we did not show the fluctuation-free
results with $\Delta=0$ that behave like the results with
$\Delta=0.02$ or $0.002$ till $g^2\mu t\sim 60$, and monotonically
approach zero beyond it.)

There are two interesting observations that one can notice at a
glance.  First, the choices of $\Delta=0.02$ and $\Delta=0.002$ make
only little change in the onset of the instability around
$g^2\mu t\sim 100$ where $\Pl/\Pt$ start growing.  Owing to this, we
can be so sure that our results should be robust at least on a
qualitative level regardless of our ignorance about the precise value
of $\Delta$.  Second, if $\Delta$ is less than $\sim 0.2$, we cannot
reach the complete isotropization.  This is quite unexpected:  Because
the simulation runs in the isotropic setup, the anisotropy given at
the initial time should naturally fade out if we wait for a
sufficiently long time.  This intuition is correct, but the point is
that it takes an extraordinarily long time unless $\Delta$ is such
large that it also modifies the initial energy density.  It is quite
instructive to see that \textit{the isotropization at later time is a
  very slow process} even in a non-expanding and symmetric box.

To discuss more microscopic dynamics, we shall split the time
evolution into three distinct characteristic regimes as follows.

\subsection{Temporarily and spatially oscillatory regime}
The pressure has oscillatory behavior in the earliest stage (i.e.\
$g^2\mu t\lesssim 15$ for $\Delta=0.2$ and $g^2\mu t\lesssim 50$ for
$\Delta=0.02$ as deduced from Fig.~\ref{fig:pressure}).  The so-called
glasma instability must be developing from lower to higher
longitudinal modes, but their effects are not yet appreciable in the
bulk thermodynamics at zero mode.  In the phenomenological sense, the
theoretical understanding in this regime is the most crucial issue,
while many theoretical efforts had been devoted to rather later-time
dynamics.  Hence, our central objective in this present paper is to
shed light on this oscillatory regime.

Obviously, the equation of state still has fast components in time,
which means that the derivative expansion should not work.  One cannot
therefore apply the hydrodynamic equations to describe the time
evolution yet.  Then, a natural question arises;  what about the
spatial structure?  Is it already smooth enough or linked somehow to
the rough structure in time?

\begin{figure*}
 \includegraphics[width=\textwidth]{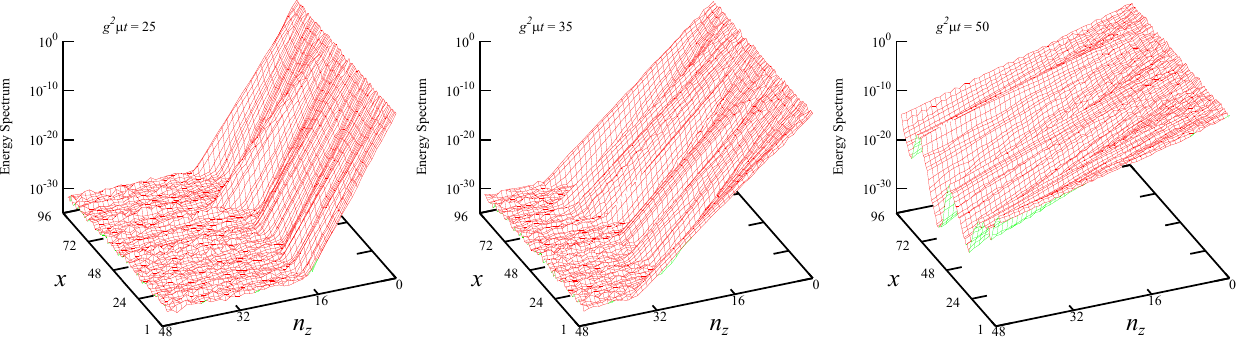}
 \caption{Energy spectrum as a function of the coordinate $x$ (in unit
   of $a$) and the momentum $k_z$ with $y=L_y/2$ fixed, calculated for
   $\Delta=0.02$.  The far left represents the result at the initial
   time $g^2\mu t=0.2$, and the time increments up to $g^2\mu t=50$ in
   the right.}
 \label{fig:avalanche}
\end{figure*}

To diagnose the microscopic dynamics, we make 3D and density plots to
illustrate the cascade flow of the energy spectrum toward higher $k_z$
or the wave number $n_z$ (where $k_z=n_z\cdot 2\pi/L_z$) in the
longitudinal direction.  We define the following energy spectrum with
only the $z$-direction Fourier transformed as
\begin{equation}
 \begin{split}
 \varepsilon(x,y,k_z) &= \sum_i 
   \tr \bigl[ E^i(x,y,-k_z) E^i(x,y,k_z) \\
 &\qquad\qquad + B^i(x,y,-k_z) B^i(x,y,k_z) \bigr]\;,
 \end{split}
\end{equation}
which is measured for each configuration.
This is not a gauge invariant quantity and we need to fix the gauge to
\textit{define} it uniquely.  We already chose the temporal axial
gauge but time-independent gauge rotations are still redundant, which
does not change the gauge-invariant observable but modifies
$\varepsilon(x,y,k_z)$.  Therefore, we fix the initial gauge
configurations~\eqref{eq:ini1} and \eqref{eq:ini2} (with fluctuations
included) to satisfy the Coulomb-gauge condition
$\boldsymbol{\nabla}\cdot\boldsymbol{A}=0$ that would flatten spiky
textures.  We used the over-relaxation method with 1000 steps to impose
the Coulomb gauge and explicitly checked that the gauge configurations
become extremely smooth then.

For qualitative discussions we may choose any $\Delta$, in principle,
but to remove an impression that our findings come from artificially
large $\Delta$, we here adopt a rather safer choice of $\Delta=0.02$
that has no effect on the very early dynamics as is clear from
Fig.~\ref{fig:pressure}.  Then, for graphical purpose, we pick a slice
of $y=L_y/2$ up and make a 3D plot of $\varepsilon(x,y,k_z)$ as a
function of $x$ and $n_z$ in Fig.~\ref{fig:avalanche} for \textit{one}
configuration.

We can understand from Fig.~\ref{fig:avalanche} what is actually
happening on the microscopic level during the oscillatory regime.  In
this very first stage the energy amplitude spreads toward larger $k_z$
triggered by spots localized in $x$ (and $y$) space.  These localized
spots look like narrow avalanches.  Let us here ``define'' what we
really mean by \textit{avalanche} for clarify.

\begin{figure}
 \includegraphics[width=\columnwidth]{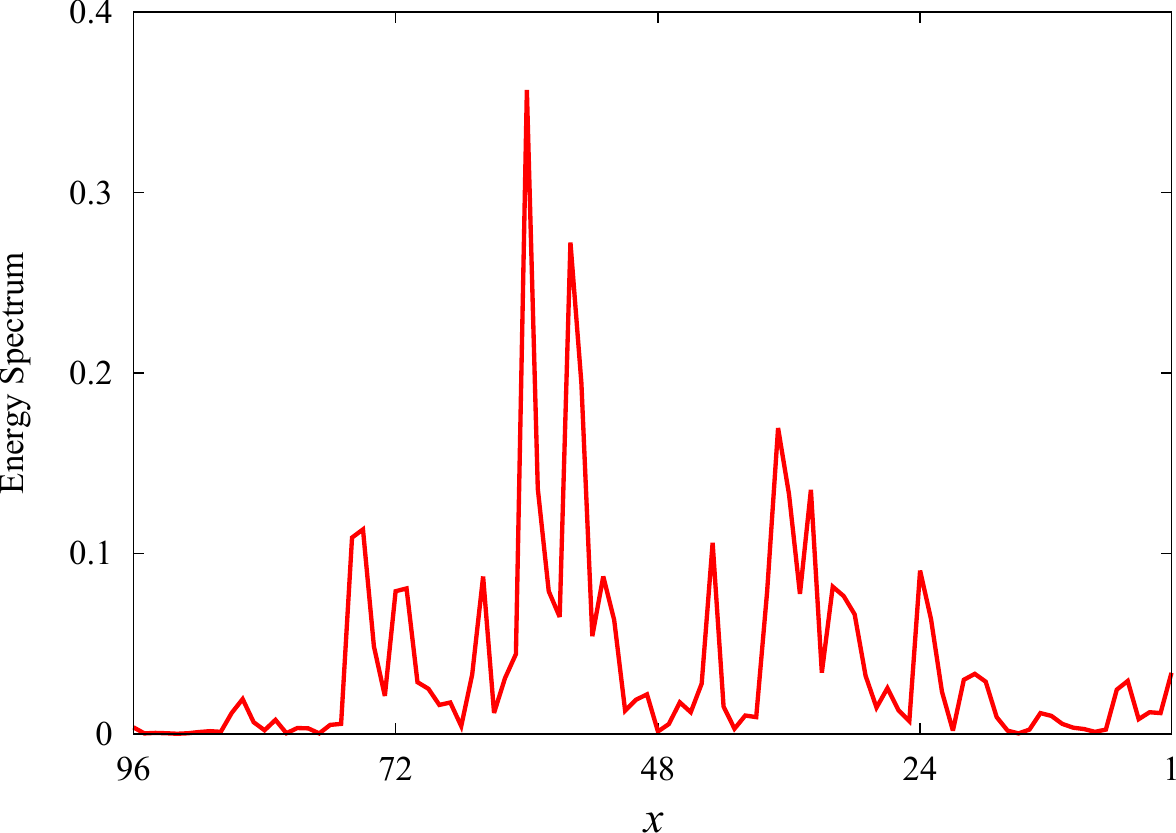}
 \caption{Energy spectrum corresponding to Fig.~\ref{fig:avalanche}
   for $\Delta=0.02$ at $n_z=0$ at the initial time $g^2\mu t=0.2$
   shown in the linear scale.}
 \label{fig:init}
\end{figure}

To do so, we have to magnify the initial energy stored at $n_z=0$
mode, which is presented in Fig.~\ref{fig:init}.  It should be noted
that this is nothing but the energy distribution already shown in the
very left of Fig.~\ref{fig:avalanche} in a form of the logarithmic
plot.  Even though Fig.~\ref{fig:init} may look like having a rough
structure, the energy fluctuates within only one order of magnitude.
It is evident from the right of Fig.~\ref{fig:avalanche} that the
inhomogeneity developing later at larger $n_z$ is correlated to the
initial pattern, and the resulting intensity differs by more than ten
orders of magnitude!  We would call this huge (but relative)
amplification of the spatial pattern the \textit{avalanche}
phenomenon.

This type of the avalanche phenomenon is quite common in many physics
problems.  The avalanche breakdown of insulator or semi-conductor is
one familiar example in which free electrons trigger the creation of
electron-hole pair.  In the present glasma simulation, we have
specified both the initial conditions~\eqref{eq:ini1},
\eqref{eq:ini2}, and the initial fluctuations~\eqref{eq:fluc}
according to the Gaussian distribution, and some local positions
happen to have irregular amplitudes as observed in
Fig.~\ref{fig:init}, which is responsible for the avalanches.
Probably they have much to do with the magnetic vortices discovered
recently in the same model setup~\cite{Dumitru:2013koh}.  We here
point out that these narrow avalanches are collective consequences
from the simultaneous existence of the glasma fields and the
fluctuation fields.  We turned the glasma background fields off as a
test calculation, and we found that the amplitudes just smoothly and
slowly decayed into higher $k_z$, but no rapid narrow avalanche
emerged.

\begin{figure*}
 \includegraphics[width=0.8\textwidth]{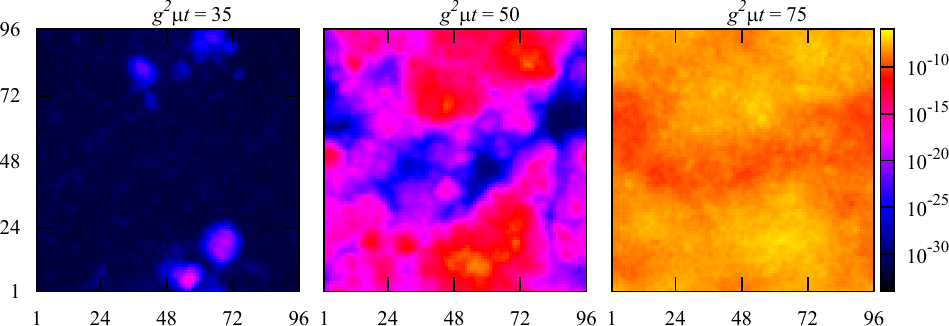}
 \caption{Snapshots of the energy amplitude for $\Delta=0.02$ at
   maximum $k_z$ on the transverse $x$-$y$ plane (in unit of $a$).
   These are taken at the times corresponding to
   Fig.~\ref{fig:avalanche}.  Clearly a dynamical pattern is formed
   first and the region with high $k_z$ next spreads over the
   transverse directions.}
 \label{fig:snap}
\end{figure*}

These avalanche-like structures gradually spread over $x$ (and $y$)
space as the time elapses, and eventually the distribution appears
uniform in transverse space at further later time, which we call the
\textit{transverse diffusion}.  To access the full transverse
structure and visualize the diffusion, we show
$\varepsilon(x,y,k_z)$ as snapshots in Fig.~\ref{fig:snap} at
$g^2\mu t=35$, $50$, and $g^2\mu t=75$ using the same configuration to
draw Fig.~\ref{fig:avalanche}.  We can then clearly perceive the
dynamical pattern formation in transverse geometry and subsequent
diffusion;  at localized spots with brighter colors we have larger
amplitudes for the maximum $k_z$ mode, namely,
$k_z^{\text{(max)}}=L_z/2 \cdot 2\pi/L_z =\pi/2$ (in unit of
$a^{-1}$).  To the best of our knowledge the present analysis is the
very first demonstration that has revealed the spontaneous generation
of spatial pattern in the real-time simulation of the Yang-Mills
theory.  At the same time as we emphasize the novelty, we would also
like to draw attention to the similarity to many other systems out of
equilibrium:  One intuitive example lies in the formation of the
magnetized domains described by the time-dependent Ginzburg-Landau
theory of the classical spin models.  Such a pattern has been
numerically discovered in the direction from the random initial state
to the ordered spin state at lower $T$, and amazingly also in the
opposite direction from the enforced ordered (or coherent) state to
the disordered (or decoherent) spin state at high $T$~\cite{Kudo}.
Our results are reminiscent of the latter associated with
non-equilibrium decohering processes.

\begin{figure}
 \includegraphics[width=\columnwidth]{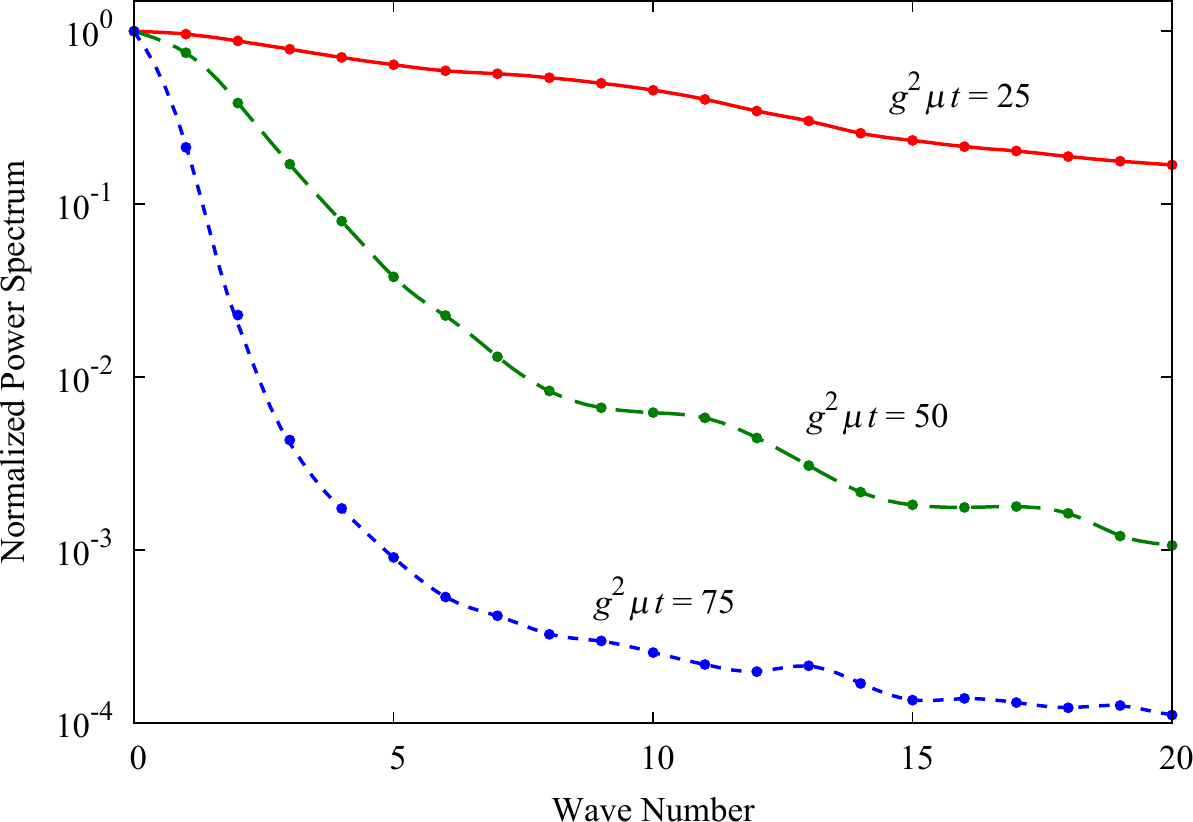}
 \caption{Power spectrum $P(k_\perp)$ as a function of the transverse
   wave-number, normalized by the zero-mode value.  The ensemble
   average is taken over 50 configurations.}
 \label{fig:power}
\end{figure}

Let us quantify the transverse diffusion processes by calculating the
energy-density correlation function from Fig.~\ref{fig:snap} or the
``power spectrum'' defined by
\begin{equation}
 P(k_\perp) = \langle\varepsilon(-k_\perp,k_z^{\text{(max)}})
  \varepsilon(k_\perp,k_z^{\text{(max)}})\rangle\;,
\end{equation}
where we note that the meaning of $k_\perp$ is totally different from
$k_z$;  we introduced $k_\perp$ as a Fourier transform of $x$ and $y$
of $\varepsilon(x,y,k_z)$, while $k_z$ refers to the momentum carried
by the chromo-electric and chromo-magnetic fields.

To absorb orders of magnitude difference at different time, we
normalize the power spectrum by the zero-mode value $P(0)$ and draw
Fig.~\ref{fig:power} for $g^2 \mu t=25$, $50$, and $75$.

We can clearly confirm that the long-range correlation becomes more
and more enhanced as the time goes, which is quite consistent with
what we can see from Fig.~\ref{fig:snap}.  (We note that the
wave-number, say 10 on this plot, corresponds to the physical scale,
$10\cdot 2\pi/L_\perp\sim 1\GeV$.)  The reason why the normalized
power spectrum seemingly looks more suppressed at larger
$g^2\mu t$ is that the zero-mode grows larger.  Thus the relative
height decreases, though the absolute height is much larger at later
time.  It should be clearly noted here that we should \textit{not}
take this enhancement of the long-range correlation for a signal of
the Bose-Einstein condensate speculated as in \cite{Blaizot:2011xf}.
We are now looking at not the particle distribution but the
energy-density correlation.  We observe that the spots in the
transverse plane spread out quickly toward uniformity, which is to be
interpreted as the diffusion as can be inferred from
Fig.~\ref{fig:snap}.

\subsection{Intermediate fast-growing regime}
After some time ($15\lesssim g^2\mu t\lesssim 30$ for $\Delta=0.2$ and
$50\lesssim g^2\mu t\lesssim 100$ for $\Delta=0.02$ in
Fig.~\ref{fig:pressure}) the equation of state behaves smoothly enough
in time and also in space, which should enable the hydrodynamic
evolution to work fine during this regime since the derivative
expansion makes sense.  The system still goes on approaching
isotropization, and thus it is neither isotropic nor thermalized yet.
An important question is what determines the typical time scale for
the transition from the oscillatory regime to the growing regime.  In
other words, we need to know what is still missing to accelerate the
onset time for the hydrodynamic evolution.  The time scale is
characterized by the balance between the initial energy density stored
at the zero mode and the rate of the energy flow that is intrinsically
determined by the Yang-Mills interactions.  Within the present
framework it is difficult to yield the onset time around a few times
$g^2\mu\sim 2\GeV\sim 0.1\;\text{fm}/c$ as required by the analysis
of the experimental data.

It is important not to be confused with the behavior of the most
unstable mode in the expanding
case~\cite{Romatschke:2005pm,Fukushima:2011nq} that looks very similar
to Fig.~\ref{fig:pressure}.  In the expanding case the onset is delayed
simply by the kinematical reason~\cite{Fujii:2008dd}, and in the
non-expanding case in Fig.~\ref{fig:pressure} it is not the most
unstable component but the whole pressure that we are dealing with.
Therefore it takes time for the instability to spread over the whole
phase space.

\subsection{Asymptotic slowly-growing regime}
The diffusion is caused by inhomogeneity, and so it becomes slower and
slower with less and less inhomogeneity and anisotropy.  Naturally the
tendency toward isotropization becomes weakened as $\Pl$ and $\Pt$ get
closer to each other.  In such an asymptotic regime
($g^2\mu t\gtrsim 30$ for $\Delta=0.2$ and $g^2\mu t\gtrsim 100$ for
$\Delta=0.02$ in Fig.~\ref{fig:pressure}) the characteristic time
scale is unphysically long.  From the experimental point of view, all
theoretical considerations in this late regime are irrelevant to the
thermalization problem.  Nevertheless, as a rather ``academic'' problem
to investigate the non-linear dynamics of the Yang-Mills theory, it is
worth paying our attention to microscopic details in this asymptotic
regime.

\begin{figure}
 \includegraphics[width=\columnwidth]{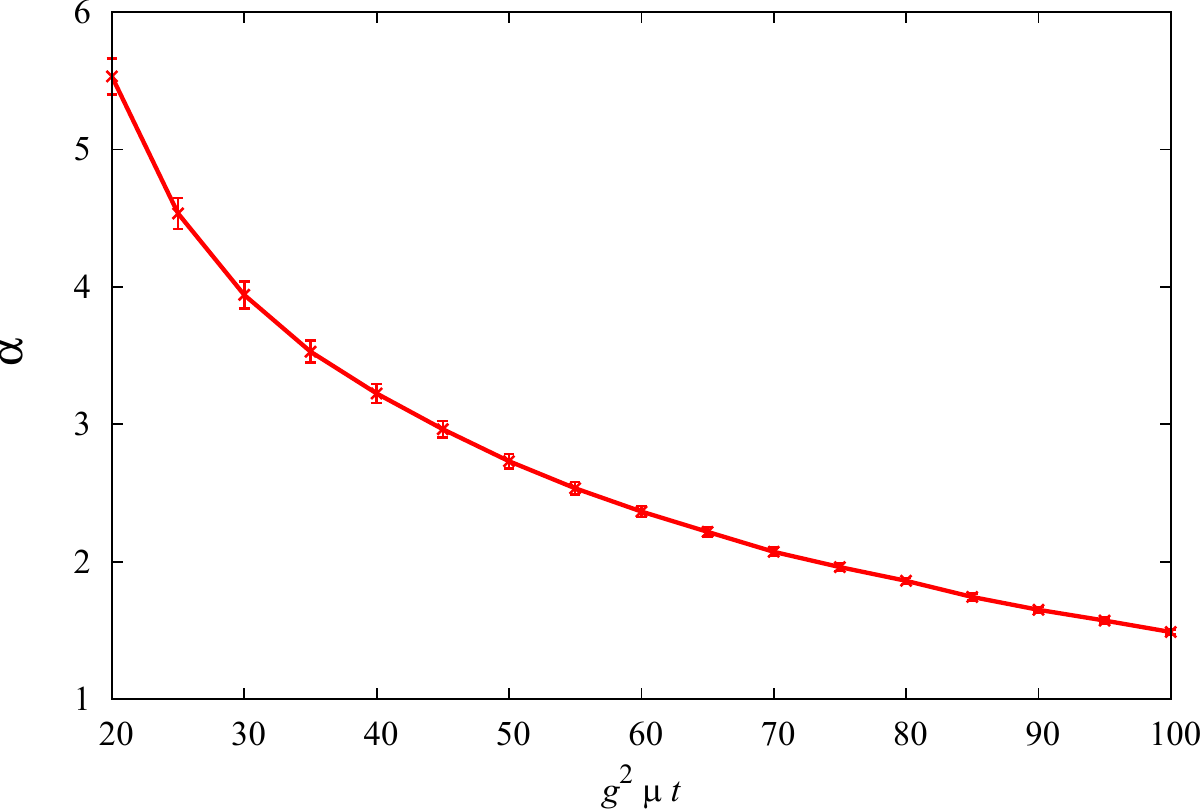}
 \caption{Power index deduced from the tail of the energy spectrum
   fitted by $k_z^{-\alpha}$ given as a function of dimensionless
   time.}
 \label{fig:scaling}
\end{figure}

Usually some kind of scaling law may be observed in a well-developed
turbulent system at late time.  In the present simulation with
specific initial conditions~\eqref{eq:ini1} and \eqref{eq:ini2},
however, the zero mode cannot be a consistent source to supply the
energy injection and so it cannot sustain a steady inertial region in
the energy spectrum.  Still, there may be a chance to see scaling
behavior at the tail of the energy spectrum.  To test this idea, we
attempt to fit the longitudinal energy spectrum by the power-law
spectrum $\sim k_z^{-\alpha}$, and we find that the fit works well in
the range, $n_z=25\sim 48$.  Then, the power $\alpha$ turns out to be
a function of time as in Ref.~\cite{Schlichting:2012es}, which is
plotted in Fig.~\ref{fig:scaling}.  The value of the index $\alpha$
decreases with increasing time, which crosses the Kolmogorov value
$5/3=1.67$ and becomes even smaller.  As we discussed above, the
inertial region is not stable and the precise value of $\alpha$ is not
very important in the present case but this level of qualitative
agreement is quite suggestive.  One might care about the consistency
with Ref.~\cite{Berges:2008mr} in which a stable power-law has been
identified.  We note that this difference between the present analysis
and Ref.~\cite{Berges:2008mr} comes from the totally different choice
of the initial conditions~\eqref{eq:ini1} and \eqref{eq:ini2} that
resemble the anisotropy in the heavy-ion collision.


\begin{figure}
 \includegraphics[width=\columnwidth]{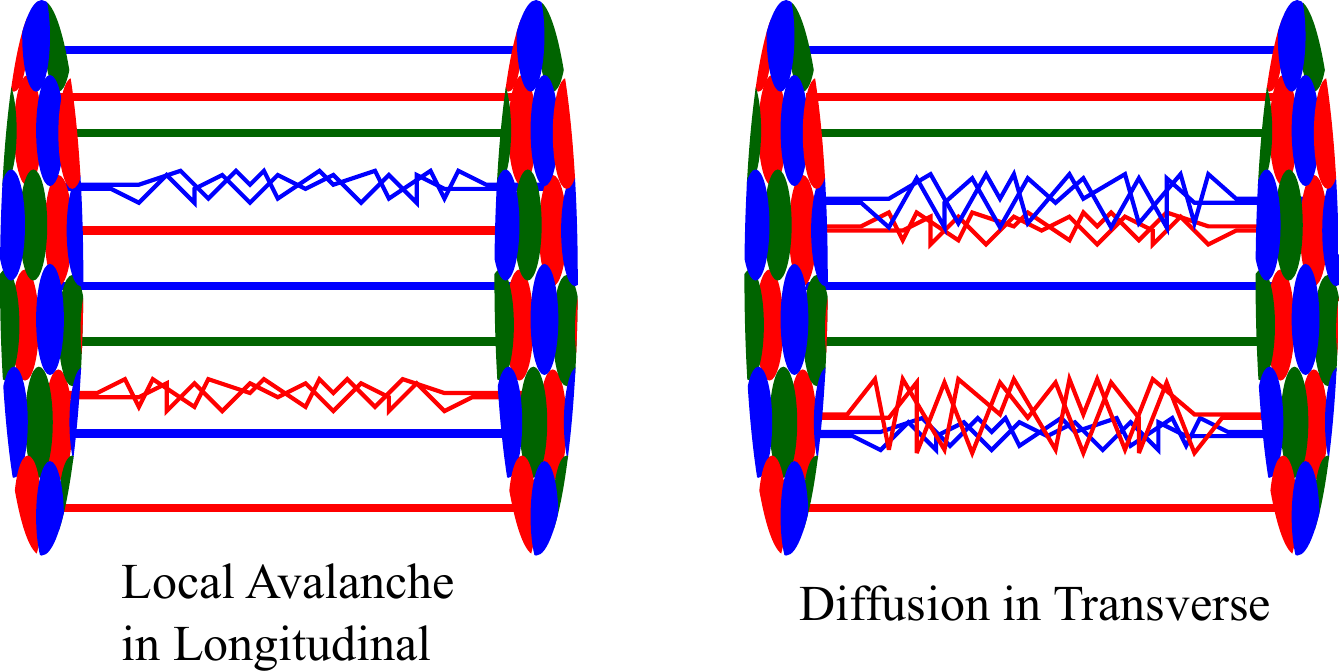}
 \caption{Schematic picture of two fastest processes in the early-time
 dynamics in the relativistic heavy-ion collision.}
 \label{fig:schematic}
\end{figure}

\section{Conclusions}
With the results from our numerical simulations, we can arrive at the
following picture of the very early-time stage in the relativistic
heavy-ion collision, as is sketched in the illustration of
Fig.~\ref{fig:schematic}.

The fastest process is driven by the avalanche-like decay along the
longitudinal direction which takes place locally in transverse plane.
These avalanches are to be attributed to initial fluctuations.  Once
this occurs, the boost invariance or the $z$-invariance is quickly but
only locally broken as in the left of Fig.~\ref{fig:schematic}.  This
view also invokes the famous Reynolds' experiment of the turbulent
flow inside of a pipe~\cite{reynolds} where the translationally steady
flow of ink begins wandering under disturbances if Reynolds' number
exceeds a critical point.  From this analogy it may well be reasonable
to identify these local avalanches as appearance of a sort of fluid
turbulence.  Also, it would be conceivable to associate them with the
QCD string breaking which is accompanied by the particle production.

The next vital fast process is the diffusion over the transverse
plane.  This turbulent diffusion is a quite efficient mechanism to
dispose energy in the whole phase space, and eventually to let the
equation of state behave smoothly enough.

Before addressing the possible relevance to the experimental data,
the following upgrades should be taken into account:  First, it is
necessary to incorporate the full quantum spectrum that should further
accelerate the process speed.  Second, related to this, we should
carefully deal with the renormalization and subtract the UV divergence
originating from the quantum fluctuation.  Third, we need to turn the
expansion on, which makes it even more subtle to handle the first and
the second points above.  In principle, as we commented, the
avalanches should be associated with the particle production, which is
to be reflected in the moments of the angular distribution of the
produced particles.  For quantitative theoretical prediction, however,
we must tackle the above-mentioned tough obstacles and complete the
thermalization scenario first.  We believe that the qualitative
finding reported in this work should be a crucial step toward solving
the puzzle of the thermalization problem.

\acknowledgments
  K.~F.\ thanks Guy Moore for comments on the gauge fixing,
  Thomas Epelbaum for useful communications, and
  Maximilian Attems, Francois Gelis, Yoshimasa Hidaka, and Keiji Saito
  for discussions.
  He was supported by JSPS KAKENHI Grant \# 24740169.

\end{document}